\documentclass[acmlarge]{acmart}
\usepackage{booktabs} 
\usepackage{natbib}
\usepackage{algpseudocode}
\usepackage{algorithm}
\usepackage{empheq}
\usepackage{graphicx}
\usepackage{tabularx}
\usepackage{multirow}
\usepackage{float}

\usepackage{subfig}
\usepackage{float}
\usepackage{caption}
\usepackage{algorithm}
\usepackage{algpseudocode}
\usepackage{color}
\usepackage{colortbl}
\usepackage{tikz}
\usepackage{multirow}

\setcopyright{rightsretained}

\begin{document}
\title{Vulnerability Coverage as an Adequacy Testing Criterion}

\author{Shuvalaxmi Dass}
\orcid{0000-0002-2469-5033}
\affiliation{%
  \institution{Texas Tech University}
  \streetaddress{P.O. Box 43104}
  \city{Lubbock}
  \state{Texas}
  \country{USA}
  \postcode{79409-3104}
}
\email{shuva93.dass@ttu.edu}

\author{Akbar Siami Namin}
\affiliation{%
  \institution{Texas Tech University}
  \streetaddress{P.O. Box 43104}
  \city{Lubbock}
  \state{Texas}
  \country{USA}
  \postcode{79409-3104}
}
\email{akbar.namin@ttu.edu}

\begin{abstract}
Mainstream software applications and tools are the configurable platforms with an enormous number of parameters along with their values. Certain settings and possible interactions between these parameters may harden (or soften) the security and robustness of these applications against some known vulnerabilities. However, the large number of vulnerabilities reported and associated with these tools make the exhaustive testing of these tools infeasible against these vulnerabilities infeasible. As an instance of general software testing problem, the research question to address is whether the system under test is robust and secure against these vulnerabilities. This paper introduces the idea of ``{\it vulnerability coverage},'' a concept to adequately test a given application for a certain classes of vulnerabilities, as reported by the National Vulnerability Database (NVD). The deriving idea is to utilize the Common Vulnerability Scoring System (CVSS) as a means to measure the fitness of test inputs generated by evolutionary algorithms and then through pattern matching identify vulnerabilities that match the generated vulnerability vectors and then test the system under test for those identified vulnerabilities. We report the performance of two evolutionary algorithms (i.e., Genetic Algorithms and Particle Swarm Optimization) in generating the vulnerability pattern vectors.      
\end{abstract}

%
%

\copyrightyear{2020}
\acmYear{2020}
\acmConference[SAC '20]{The 35th ACM/SIGAPP Symposium on Applied Computing}{March 30-April 3, 2020}{Brno, Czech Republic}
\acmBooktitle{The 35th ACM/SIGAPP Symposium on Applied Computing (SAC '20), March 30-April 3, 2020, Brno, Czech Republic}\acmDOI{10.1145/3341105.3374099}
\acmISBN{978-1-4503-6866-7/20/03}



\maketitle

\section{Introduction}
\label{sec:introduction}

Software systems and applications are often released with a great number of features and settings. These features and configurations serve their users and the underlying platforms for different purposes such as architectural settings, virtualization, performance, security and access control, privacy, and system level interactions. For instance, MySQL Version 5.5 lists more than 600 configuration parameters categorized into 3  groups namely Server Options, System Variables, and Status Variable References \cite{Mysql}. While these parameters offer great features to their administrators for setting up software systems properly, an improper configuration and setting of such parameters also create loopholes in the systems and thus are vulnerable to certain known or even unknown security attacks (i.e., zero-day vulnerability \cite{DBLP:conf/bigdata/Abri}). 

According to the National Vulnerability Database (NVD) \cite{NVD}, as of September 2019, there are $1,644$ records of reported vulnerabilities with assigned CVE numbers. Some of these vulnerabilities are directly the cause of improper settings of the configurations parameters offered as features by the software systems. From the software testing perspective, enumerating all configuration settings and then verifying whether the given software is vulnerable to certain attacks is infeasible. 

This paper introduces the concept of ``{\it vulnerability coverage}'' as an adequacy criterion for choosing instances of vulnerabilities that the software under test needs to be checked against. The deriving idea is to utilize the Common Vulnerability Scoring System (CVSS) as a means to measure the vulnerability level of the software under test. For instance, a vulnerability  with CVE-2019-16383 reported for MySQL has the severity rated as $8.2$ out of 10 and it is labeled as high. 
In this paper, we explore vulnerability coverage, as an adequacy criterion for choosing the vulnerabilities needed to be examined for the software under test. The key contributions of this work are as the following:
\begin{enumerate}
    \item Introduce the novel idea of vulnerability coverage to be utilized as an adequacy criterion for testing systems and their configurations thoroughly. 
    \item The evolutionary algorithms (i.e., genetic algorithms (GA) and particle swarm optimisations (PSO)) are adapted to generate adequate test inputs with respect to the introduced vulnerability coverage criterion. 
    \item The performance of the proposed vulnerability coverage criterion is reported through a case study demonstrating the feasibility of the proposed coverage criterion. 
\end{enumerate}
We will be using the words \textit{Vulnerability Vector/Pattern}, \textit{CVSS vector}, \textit{configurations}, interchangeably. This paper is structured as follows: Section \ref{sec:relatedwork} reviews the related works. The concept of Common Vulnerability Scoring System (CVSS) is presented in Section \ref{sec:CVSS}. The methodology of adapting the evolutionary algorithms is presented in Section \ref{sec:methodolgy}. The evaluation of the proposed idea performed on a case study is reported in Section \ref{sec:evaluation}. 
Section \ref{sec:conclusion} concludes the paper.


\section{Related Work}
\label{sec:relatedwork}

The work presented in this paper offers some solutions for generating a set of thorough secure configurations for a given system when implementing Moving Target Defense (MTD) strategies \cite{DBLP:journals/jcst/ZhengN19}.

Crouse and Fulp \cite{6111663} used conventional Genetic Algorithms to implement a Moving Target Defense (MTD) environment to enable security through temporal and spatial diversity in computer configuration parameters.
 Crouse and Fulp 
 reported that the pool of configurations becomes stale when there are no changes introduced to the set of configurations over a period of time. As a result, GA deals with a limited set of configurations.
In this approach, the fitness (security) of  aging configurations is reduced by a value (i.e., decay value)  based on the time since they were last active. Such weak configurations are eventually replaced by more secure ones.

Lucas et al.\ \cite{Lucas:2014:IFE:2602087.2602100} described a framework for implementing MTD at host-level. Their framework uses evolutionary-inspired  Genetic Algorithm (GA) to generate secure configurations. 
They evaluated their framework using two qualitative measurements: Fitness score and pairwise Hamming distance (i.e., diversity).

The use of genetic algorithms in generating a thorough set of test inputs has been discussed and modeled in literature. For instance, Andrews et al.\ \cite{DBLP:conf/kbse/AndrewsLM07} used genetic algorithms to enable random testing more effective. A similar approach is adapted here to produce a better test inputs for the purpose of maximizing the coverage level of test pool using the evolutionary algorithms.

\section{Common Vulnerability Scoring}
\label{sec:CVSS}

The Common Vulnerability Scoring System (CVSS) provides a way to capture the principal characteristics of a vulnerability and produce a numerical score reflecting its severity. The scoring system also provides a textual representation of the semantic of the calculated score. The numerical score can then be translated into a qualitative representation (e.g., low, medium, high, and critical) to help organizations to properly assess and prioritize their vulnerability management processes \cite{first}.

CVSS is composed of three main metric groups: (1) Base, (2) Temporal, and (3) Environmental, each consisting of a set of sub-metrics. Without loss of generality and to demonstrate the feasibility of the proposed approach, the GA and PSO algorithms are only applied to the Base metric group. Additional reason that this paper focuses on the Base metric is due to the fact that the Base metric quantifies the essential characteristics of a vulnerability, which remains unchanged across different environments and over time. The Base metric consists of two sub-main metrics:

\textbf{(1) Exploitability Metrics.}
     It describes the ``how'' part of the attack that is being captured, which depends on the characteristics of the vulnerable components. This metric consists of:
    \begin{itemize}
        \item[--] \textit{Attack Vector (AV).} It reflects the proximity of the attacker to attack the vulnerable component. The more the proximity required to attack the component, the harder it is for the attacker. The attack vector takes on four values: Network
        (N), Adjacent (A), Local (L) and Physical (P).
        
        \item[--] \textit{Attack Complexity (AC).} This metric reflects the resources and conditions that are required to conduct the exploit on the vulnerable component. The more the number of conditions to be met, the higher the degree of complexity of attack is. It takes on two values: Low (L), and High (H).
        
        \item[--] \textit{Privileges Required (PR).} This metric represents the level of privileges required by an attacker to successfully launch an exploit. The lesser the level is, the easier the attack is. It takes on three values:
        None (N), Low (L), and High (H).
        
        \item[--] \textit{User Interaction (UI).} It reflects whether the participation of the user is required for launching a successful attack. The attack becomes difficult if the user interaction is mandatory. This metric takes on two values: None (N) and Required (R).  
        
    \end{itemize}

\textbf{(2) Impact Metrics.} These metrics reflect the characteristics of the impacted components. They consist of:
\begin{itemize}
    \item[--] \textit{Availability Impact.} It measures the severity of the attack on the availability of the impacted component. The metric takes on these values:  None (N), Low (L), and High (H).
    \item[--] \textit{Integrity Impact.} It measures the severity of the attack on the integrity of the impacted component. 
    The metric takes on three values:  None (N), Low (L), and High (H).
    \item[--] \textit{Confidentiality Impact.} It measures the severity of the attack on the confidentiality of the impacted component. It takes on the following values: None (N), Low (L), and High (H).

\end{itemize}

\textbf{(3) Scope Metrics.} There is also a vector called {\it ``scope''} which describes the scope of the attack (i.e., whether the attack on the vulnerable component consequently impacted the resources beyond its means). It takes on two values: Unchanged (U) and Changed (C).

\begin{figure}
  \includegraphics[width=8cm]{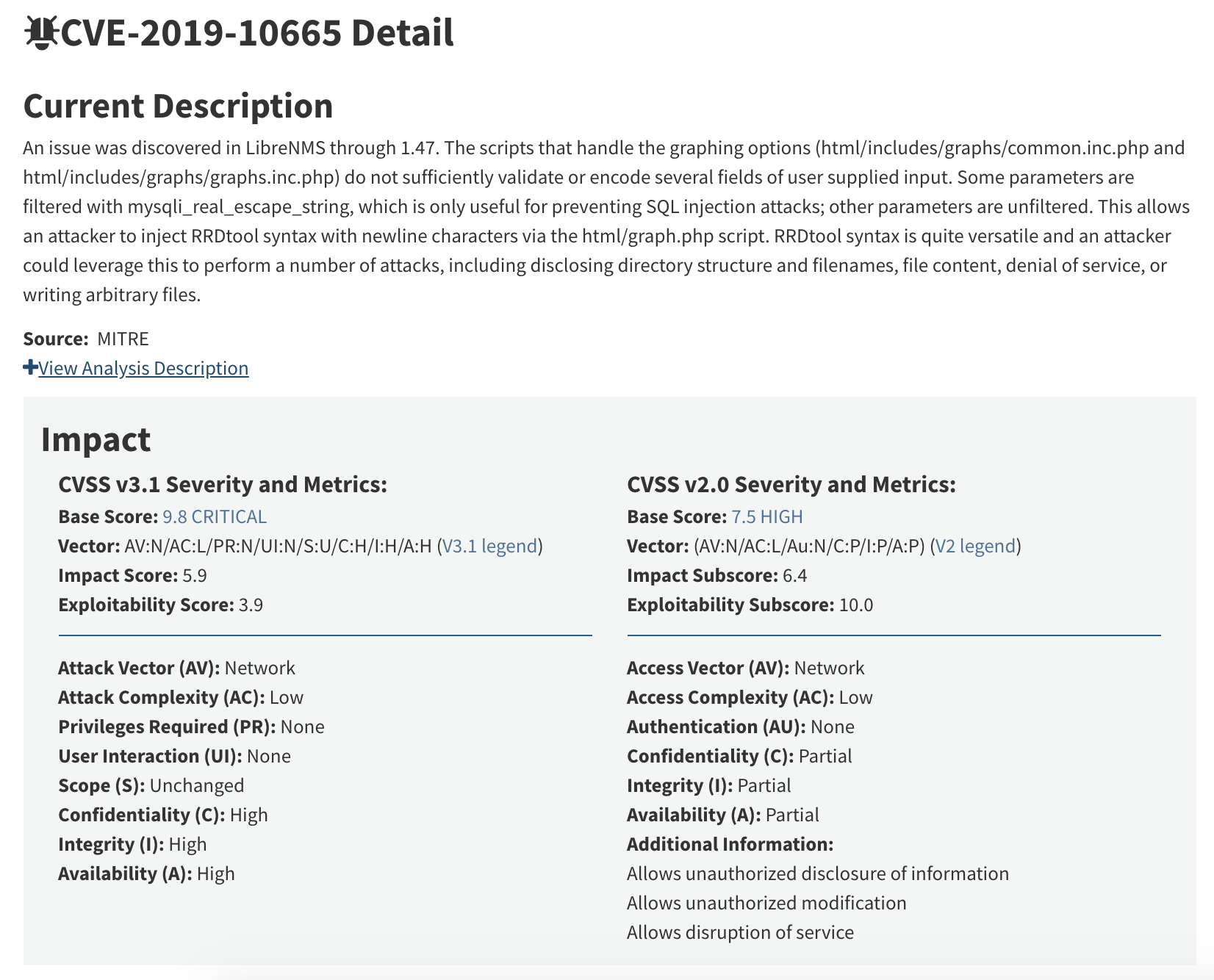}
  \caption{The description of CVE-2019-10665 for MySQL.}
  \label{fig:CVESample}
  \vspace*{-0.15in}
\end{figure}

CVSS  incorporates all of the aforementioned metrics in a formula to calculate the vulnerability score. The lower the Base score is, the harder it is for the execution of an exploit on the vulnerable component. Furthermore, the score is measured using a certain number of features. For instance, Figure \ref{fig:CVESample} shows the CVSS score along with the vulnerability vector for CVE-2019-10665 reported for MySQL. As the figure shows, the severity of this vulnerability is rated as $9.8$ out of 10 and it is labeled as a critical one. The generated format of the vulnerability/CVSS vector is {\tt [AV:N/AC:L/PR:N/UI:N/S:U/C:H/I:H/A:H]} where \cite{first} :
\begin{itemize}
    \item[--] {\tt AV:N} indicates that the Attack Vector (AV) of such vulnerability is set at the Network (N) level. 
    \item[--] {\tt AC:L} implies that the Attack Complexity (AC) of this vulnerability is Low (L).
    \item[--] {\tt PR:N} shows that the Privileges Required (PR) for launching an attack based on this vulnerability is None (N). 
    \item[--] {\tt UI:N} reports that the User Interaction (UI) and involvement in enabling launching a successful attack is None (n). 
    \item[--] {\tt S:U} shows the Scope (S) of the attack is Unchanged (U). 
    \item[--] {\tt C:H} indicates that the risk of losing the Confidentiality (C) of data when this attack occurs is High (H). 
    \item[--] {\tt I:H} implies that the risk of losing the Integrity (I) of data when this attack occurs is High (H). 
    \item[--] {\tt A:H} indicates that the risk of losing the Availability (C) of data when this attack occurs is also High (H). 
\end{itemize}



\section{Methodology}
\label{sec:methodolgy}

We compared two different evolutionary algorithms  for the vulnerability pattern generation targeting a certain level of CVSS score, as fitness function. The optimization algorithms that we implemented were: 1) Genetic and evolutionary Algorithms (GA), and 2) Particle Swarm Optimization (PSO).

We chose the CVSS score as a measure of fitness function because the diversity of its parameters enable us to adapt typical optimization and search techniques such as evolutionary algorithms for addressing this problem. Moreover, some other researchers have also adapted these types of greedy algorithms to address similar problems in test input generations in the context of random testing (e.g., \cite{DBLP:conf/kbse/AndrewsLM07}). Each parameter in the configuration is assigned a score vector that represents the vulnerability it contains along with its severity. The score is modeled after the Common Vulnerability Scoring System (CVSS) vector and provides a method for measuring the security level of an individual configuration parameter setting. The vector can serve as a foundation to estimate the number of possible vulnerabilities of a certain configuration. Moreover, to account for the diversity in the configuration generation, we also calculated the Hamming distance, which measures how different one pattern (i.e., configuration) is from another. 

\subsection{Genetic Algorithm}
We developed a python-based genetic algorithm script  to generate a CVSS vector string pool with the best fitness score (i.e., $ = 2.0$). We call it a ``{\it string}'' since each vector is treated as a string in our implementation. We set the number of iterations $= 50$ and population size $= 100$. The algorithm mainly comprises of five parts:

\begin{enumerate}
    \item \textit{Configuration Generation}: An initial pool of 100 possible CVSS vector strings was generated randomly.
    \item \textit{Fitness Score}:  The Python library, called {\tt CVSS3}, implements the Base metric score  it was utilized to calculate the CVSS scores. The vector strings were assigned their CVSS scores to be their fitness scores if the former lied between $2.0$ and $5.5$, otherwise, they were assigned $100$.
    \item \textit{Breeder's Selection}: This type of selection not only chooses the best solutions (i.e., vectors with lower score) of the previous generation but also selects some random ones to avoid converging soon to a local minima. 
    \item \textit{Crossover}: For crossover, we took the simplest way of randomly switching the values of corresponding metrics among the two parent vector strings.
    \item \textit{Mutation}: We performed mutation on the CVSS vector configurations by randomly picking one vector field and changing its value by randomly selecting from permissible values.
\end{enumerate}

\subsection{Particle Swarm Optimization Algorithm}
We also implemented the PSO algorithm based on the CVSS scores for the purpose of comparing the performance of PSO  with GA on generating a set of secure configurations. In order to enable the comparison unbiased, we kept the best score, number of iterations and the size of population similar to the ones set for GA.

The PSO implementation is similar to GA. However, unlike GA, PSO is easier to implement as it has no evolution operators such as crossover and mutation. In PSO, the potential solutions, are called particles. The algorithm mainly deals with two parameters: (1) {\tt pbest\_fitness} and (2) {\tt particle\_vel}, which contain the initial {\tt pbest} fitness and velocity values for every particle in the swarm, respectively.
 The velocity for each particle measures how far is its fitness score ({\tt pbest}) from the best score.

The PSO algorithm returns a count of particles with scores $ = 2.0$ or velocity $=0$ in every iteration. We set the target global best value as $10.0$,  {\tt particle\_velocity} (i.e., Hamming Distance) between the range $[0, 8]$ as 0 and 8 are the minimum and maximum  number of differences that might exist between two particles, respectively. The {\tt fitness\_range} is set in the range of $[2, 10]$ where $2.0$ is deemed as the best fit and $10.0$ is considered to be poorly unfit.

The \textit{Particle Initialization} follows a similar strategy as described for the configuration generation in GA. The algorithm continues to find the best fitness value and velocity for every particle until it reaches a threshold limit. In each iteration, the algorithm assigns the {\tt pbest\_fitness} (particle best) values as their CVSS scores {\tt cvss\_fit} only when the fitness is better (i.e., in our case, lesser the better) than its current {\tt pbest} fitness value. It then assigns the {\tt gbest} (global best) value of the swarm to be the best {\tt pbest} value obtained so far by any particle in the population.
After finding its two best fitness values, each particle updates itself in a similar fashion as GA where the configurations are mutated whenever their current velocity values are greater than their previous ones. For instance, if `AV' vector field gets chosen randomly, then any one value will be randomly picked from the \{H, L, N, A\} set. 
 

\section{Evaluation and Discussion}
\label{sec:evaluation}
We evaluated the performance of the two evolutionary algorithms through a case study. Table \ref{tab:PSOvsGA} reports the percentages of number of CVSS vulnerability vectors produced for different CVSS score ranges across 100 runs for both GA and PSO scripts. We observed that the values came out to be almost similar. 

\begin{table}[h]
\centering
\begin{tabular}{|l|l|l|l|l|}
\hline
\textbf{}                            & \textbf{[2.0]} & \textbf{(2.0, 3.0]} & \textbf{(2.0, 4.0]} & \textbf{(2.0, 5.0]} \\ \hline
\textbf{GA}  & 3.505         & 22.956         & 34.972         & 38.567         \\ \hline
\textbf{PSO} & 3.387         & 23.144         & 34.088         & 39.379         \\ \hline
\end{tabular}
\caption{The percentage instances of CVSS vectors produced by GA and PSO for the different score ranges in 100 runs.}
\label{tab:PSOvsGA}
  \vspace*{-0.3in}
\end{table}

\begin{table*}[!t]
\centering
\begin{tabular}{|c|c|c|m{11.5cm}|c|}
\hline
\multicolumn{1}{|c|}{\bf \#} & \multicolumn{1}{c|}{\bf VENDOR} & \multicolumn{1}{c|}{\bf CVE  ID} & \multicolumn{1}{c|}{\bf VULNERABILITY DESCRIPTION}   & 
\multicolumn{1}{c|}{\bf Score} \\
\hline
\textbf{1} & Mysql    & CVE-2019-14939   & An issue was discovered in the mysql (aka mysqljs) module 2.17.1 for Node.js. The LOAD DATA LOCAL INFILE option is open by default  & 2.1     \\ \hline
\textbf{2} & Mysql           & CVE-2016-7440    & The C software implementation of AES Encryption and Decryption in $wolfSSL$ (formerly CyaSSL) before 3.9.10 makes it easier for local users to discover AES keys by leveraging cache-bank timing differences                                      & 2.1            \\ \hline
\textbf{3} & Oracle          & CVE-2014-6551    & Unspecified vulnerability in Oracle MySQL Server 5.5.38 and earlier and 5.6.19 and earlier allows local users to affect confidentiality via vectors related to CLIENT:MYSQLADMIN                                                                & 2.1            \\ \hline
\textbf{4} & Oracle          & CVE-2012-3160    & Unspecified vulnerability in the MySQL Server component in Oracle MySQL 5.1.65 and earlier, and 5.5.27 and earlier, allows local users to affect confidentiality via unknown vectors related to Server Installation                             & 2.1            \\ \hline
\textbf{5} & Mysql           & CVE-2006-4031    & MySQL 4.1 before 4.1.21 and 5.0 before 5.0.24 allows a local user to access a table through a previously created MERGE table, even after the user's privileges are revoked for the original table, which might violate intended security policy & 2.1            \\ \hline
\end{tabular}
\caption{The number of vulnerabilities found for CVSS pattern: {\tt AV:L/AC:L/PR:N/S:N/C:P/I:N/A:N}.}
\label{tab:my-table}
\vspace{-0.35in}
\end{table*}

A possible explanation of obtaining similar results for both GA and PSO is that the number of vector fields for CVSS (i.e., Base level) is limited to eight. As a result, there are not too many options for GA or PSO to select from. Therefore, both algorithms converge to the same values quickly because the search space is very small. Considering all the other metrics in CVSS might provide a larger search space for the algorithms. 



To illustrate the effectiveness of the introduced adequacy criterion for security testing, we looked up the CVE website \cite{cve} and identified different vulnerabilities whose CVSS patterns matched the vulnerability pattern produced by GA and PSO. As an example, based on the CVSS pattern: {\tt AV:L/AC:L/PR:N/S:N/C:P/I:N/A:N} (In C: P , P stands for Partial or Low as per the website), table \ref{tab:my-table} shows multiple CVEs of vulnerabilities we found in the product MySQL for different vendors along with their description and CVSS score with the exact CVSS vector pattern matching.

\section{Conclusion and Future Work}
\label{sec:conclusion}

This paper introduced the concept of ``{\tt vulnerability coverage}'' as an adequacy criterion for security and vulnerability testing of software applications. The deriving idea is to utilize Common Vulnerability Scoring System (CVSS) as a fitness metric and identify a set of vulnerability vector patterns that achieves a certain level of CVSS score. The generated set can be used for adequacy testing of underlying system in which all or representative sets of vulnerabilities with similar vulnerability vector patterns will be selected for further inspection of the system under test. The paper compared two evolutionary-based algorithms namely Genetic Algorithms and Participle Swarm Optimization and the results indicated a similar results obtained by these two greedy algorithms. 

The novel idea of vulnerability adequacy criterion as introduced in this paper needs further attentions. To our best knowledge, an adequacy criterion based on vulnerability coverage does not exist. There are several other features that need to be investigated including other metrics incorporated into CVSS and National Vulnerability Database (NVD) including temporal and environmental metrics. Furthermore, the idea needs tool supports and further empirical studies to thoroughly search the NVD database for reported vulnerabilities with exact pattern matching property for security testing purposes and investigate the effectiveness of such adequacy criterion.

The usefulness of Bayesian approaches have been discussed extensively in the literature \cite{DBLP:conf/sigsoft/NaminS10}. These probabilistic reasoning approaches can be adapted in the context of uncertainty analysis for implementing adaptive security testing in dynamic domains (e.g., reinforcement reasoning \cite{DBLP:conf/compsac/ChatterjeeN19}).
It is possible to apply these learning-based algorithms along  with temporal properties and dependencies and then adapt deep learning-based approaches \cite{DBLP:conf/bigdata/Siami-Namini} to address the problem. In the presence of existence of some constraints in the configuration settings, the problem can be formulated as a constraint satisfaction problem and the generation of test inputs using symbolic executions \cite{DBLP:conf/sac/HeimlichN19}. 

\section*{Acknowledgment}
\label{sec:conclusion}
This research work is supported in part by a funding from National Science Foundation under grant numbers 1516636 and 1821560.

\bibliographystyle{ACM-Reference-Format}

\bibliography{References} 

\end{document}